\documentclass{aa}
\usepackage[dvips]{graphicx}
\usepackage{natbib}
\bibpunct{(}{)}{;}{a}{}{,}
\usepackage[mathscr]{eucal}
\begin{document}
\title{Dust shell model of the water fountain source IRAS 16342--3814}

\author{
	K. Murakawa\inst{1}
	and
        H. Izumiura\inst{2}
	}


\institute{
$^{1}$School of Physics and Astronomy, EC Stoner Building, University of Leeds,
Leeds LS2 9JT, United Kingdom\\
$^{2}$Okayama Astrophysical Observatory, 3037-5 Honjo, Kamogata, Asakuchi, Okayama, 719-0232 Japan\\
	}


\abstract{}
%
{
We investigate the circumstellar dust shell of the water fountain source
IRAS 16342--3814.
}
%
{
We performed two-dimensional radiative transfer modeling of the dust shell,
taking into account previously observed spectral energy distributions (SEDs)
and our new $J$-band imaging and $H$- and $K_S$-band imaging polarimetry
obtained using the VLT$/$NACO instrument.
}
%
{
Previous observations expect an optically thick torus in the equatorial plane
because of a striking bipolar appearance and a large viewing angle of 30 --
40$\degr$.  However, models with such a torus as well as a bipolar lobe and
an AGB shell cannot fit the SED and the images simultaneously.  We find that
an additional optically and geometrically thick disk located inside a massive
torus solves this problem.  The masses of the disk and the torus are estimated
to be 0.01~$M_{\sun}$ at the $a_\mathrm{max}=100~\mu$m dust and 1~$M_{\sun}$
at $a_\mathrm{max}=10~\mu$m dust, respectively.
}
%
{
We discuss a possible formation scenario for the disk and torus based on
a similar mechanism to the equatorial back flow.  IRAS 16342--3814 is expected
to undergo mass loss at a high rate.  The radiation from the central star is
shielded by the dust that was ejected in the subsequent mass loss event.
As a result, the radiation pressure on dust particles cannot govern the motion
of the particles anymore.  The mass loss flow can be concentrated in the
equatorial plane by help of an interaction, which might be the gravitational
attraction by the companion, if it exists in IRAS 16342--3814.  A fraction of
the ejecta is captured in a circum-companion or circum-binary disk and the
remains are escaping from the central star(s) and form the massive torus.
}


\titlerunning{dust shell model of I16342}
\authorrunning{Murakawa et al.}
\maketitle

\section{Introduction}
IRAS 16342$-$3814 (hereafter I16342), also known as OH 344.1+5.8, is an
oxygen-rich proto(pre)-planetary nebula (PPN) exhibiting a striking pair of
bipolar lobes.  \cite{he08} detected a low \element[][12]CO$/$\element[][13]CO
intensity ratio of 1.7 and concluded that \element[][12]C is being converted
into \element[][13]C via the hot bottom burning process, suggesting that the
initial mass of the central star is as high as $\ga$5~$M_{\sun}$.
Another characteristic of this object is high-velocity maser components.
The velocities of OH masers at 1612, 1665, and 1667~MHz are $\sim$65~kms$^{-1}$,
which are much faster than those in many OH$/$IR stars \citep{lm88}.
The H$_2$O masers are even faster $\sim$130 -- 180~kms$^{-1}$ \citep{zl87,lm88}.
The apparent dynamical ages of the masers are very short at $\la100$~yr
\citep{claussen09,imai12}.  The H$_2$O masers are more like jets in appearance
than the typical AGB winds.  Because of these maser properties, objects
like I16342 are called ``water fountain source (WFS)'' \citep{lm88,imai07}.
So far, 14 objects are classified into this group
\citep{imai07,suarez09,walsh09,gomez11}.  Many PNs exhibit jets, which cannot
be explained with only the generalised interacting stellar wind (GISW) model
\citep{st98}.  The WFSs are thought to be progenitors that provide crucial
clues to understanding the shaping mechanisms of these PNs.

Because of the peculiar maser properties in WFS, many works have studied
H$_2$O and OH masers.  In I16342, the best-studied object, the circumstellar
dust shell (CDS) has been studied.  The optical and near-infrared (NIR) images
show a distinct bipolar appearance with an extension of $\sim$$5\arcsec$ lying
at a position angle of $73\degr$ \citep{sahai99}.  In the optical, the eastern
lobe is fainter by a factor of $\sim$0.1 with respect to the western one.
\citet{sahai99} estimated the viewing angle to be $\sim$$40\degr$ from the 1612
MHz OH maser distribution, i.e. the western polar axis is tilted by
$\sim$$40\degr$ toward the observer.  The Keck II $L'$-band images show
a corkscrew-like feature in the bipolar lobes \citep{sahai05}.
Although a clear correlation with the distribution of the H$_2$O maser jets is
not found, the feature is probably a kind of precessing jet that carved out
the inner part of the bipolar lobe.  In the mid-infrared (MIR), an elliptic
flux structure was detected \citep{meixner99,dwk03}.  However, the
high-resolution VISIR$/$VLT data resolved a pair of bipolar lobes with
a separation of the flux peaks of 0\farcs92 at 11.85~$\mu$m
\citep{verhoelst09,lagadec11}.  These results suggest an optically and
geometrically thick torus in the equatorial plane of this object.

In this present work, we have modeled the dust shell of I16342 by means of
two-dimensional radiative transfer calculations.  The spectral energy
distribution (SED) collected from various sources and our new NIR polarimetric
data obtained using the VLT$/$NACO instrument were used to constrain
the physical parameters of the CDS.  In Sects.\,\ref{observation} and
\ref{model}, our VLT$/$NACO observations and radiative transfer calculations
are presented.  In Sect.\,\ref{discussion}, we discuss the structure and the
formation of the inner most region of the dust shell.

\section{VLT$/$NACO imaging polarimetry}\label{observation}
\subsection{Observations and data reduction}\label{reduction}
We obtained standard (non-polarimetric) images in the $J$-band and polarimetric
images in the $H$ and $K_S$ band using the CONICA camera with the NAOS adaptive
optics (AO) system mounted on the Very Large Telescope (VLT).  The S27 camera
with a pixel scale of 27.15~mas~pix$^{-1}$ was operated in the
\textsf{Double\_RdRstRd} read-out mode.  For the wavefront sensing, the N20C80
dichroic filter was used.  Because the target is extended in the optical and
near-infrared and the central star feature is invisible, a single star 2MASS
J16374030--3820087 ($m_K=8.0$ mag) at a 9\farcs7 separation from the target
was monitored for the AO reference.

The $J$-band imaging was carried out on August 4, 2009.  The natural seeing was
0\farcs5 -- 0\farcs6 according to the ESO observatories ambient conditions data
base.  The exposure time was 30 sec. per frame and six images were taken at each
four jitter positions, yielding a total integration time of 12 min.
The raw data were reduced by subtracting the dark frame, dividing by the flat
frame, and combining the all science data.  We used the photometric data of
GSPC S875-C ($m_J=11.085$ mag) for the flux calibration, which was offered in
the ESO standard calibration plan.  The signal-to-noise ratios per beam is
40 -- 240 in regions with the surface brightnesses of
$2\times10^{-15}$~Wm$^{-2}$$\mu$m$^{-1}$arcsec$^{-2}$ or brighter.

The $H$ and $K_S$ band imaging polarimetry was carried out on 2 May, 2009.
The natural seeing varied between 1\farcs0 and 1\farcs5 during the observations.
The turnable half-waveplate (HWP) and the Wollaston prism with a slit width of
$3\arcsec$ were used to measure linear polarization.  The slit position is
aligned along the long axis of the target at a position angle of 73\degr.
Four image sets were taken at position angles of the HWP of 0\degr, 45\degr,
22.5\degr, and 67.5\degr.  Seven and ten jitter positions were performed and
exposure times were 60 sec. and 30 sec. per frame, in the $H$ and $K_S$ band,
respectively.  The total integration times were 28 min. in the $H$ band and
20 min. in the $K_S$ band.  After the subtraction of the dark frames, the
sky background levels and flat fielding, the Stokes $IQU$ parameter images
were obtained by combining the ordinary and extra-ordinary ray frames.
The data of a single star HD~147283 were used for polarization calibration
\citep[$P_H=0.91\pm0.05$~\% and $P_K=0.48\pm0.04$~\%;][]{whittet92}.
Our uncalibrated measurements are $P_H=3.6$~\% and $P_K=1.7$~\%.
These discrepancies include the instrumental polarization and some other error
sources.  In observations that require a polarization accuracy better than
1~\%, such as the line spectropolarimetry, a careful calibration is essential
\citep[see][]{witzel11}.  On the other hand, an $\sim$2~\% accuracy is
acceptable in most cases of optical and infrared imaging polarimetry of
bipolar reflection nebulae.  The difference of these tolerances also results
from the different interests of the observations.  In imaging polarimetry,
the spatial information is important, but is affected by the photon noise and
the temporal variations in the shape of the point-spread function (PSF) because
of the use of small ``aperture'' sizes, e.g.\,beam sizes.  It is opposite in
the aperture polarimetry or line spectropolarimetry.  We assumed that the
discrepancies between our observations and the previous aperture polarimetry
\citep{whittet92} are caused by the instrumental polarization and the offset
Stokes $Q/I$ and $U/I$ values.  The estimated signal-to-noise ratios per beam of
the Stokes $I$ images in regions with surface brightness of
$2\times10^{-15}$~Wm$^{-2}$$\mu$m$^{-1}$arcsec$^{-2}$ or brighter are 60 -- 120
in the $H$ band and 20 -- 120 in the $K_S$ band.  The errors of linear
polarization are $P_H\sim3$~\% and $P_K\sim2$~\%.

\begin{table}
  \begin{center}
  \caption[]{Parameters of the PSF model.}
  \label{psf_parm}
  \begin{tabular}{lllll}
  \hline
    band   & FWHM     & \multicolumn{2}{c}{Moffat}  & Gaussian          \\
  \hline
           & (arcsec) & $\sigma$ (arcsec) & $\beta$ & $\sigma$ (arcsec) \\
  \hline
  $J$      & 0.11     & 0.072             & 1.28    & 0.056             \\
  $H$      & 0.12     & 0.080             & 1.07    & 0.067             \\
  $K_S$    & 0.070    & 0.043             & 1.16    & 0.031             \\
  \hline
  \end{tabular}
  \end{center}
\end{table}

\begin{figure}
  \resizebox{\hsize}{!}{\includegraphics[angle=-90]{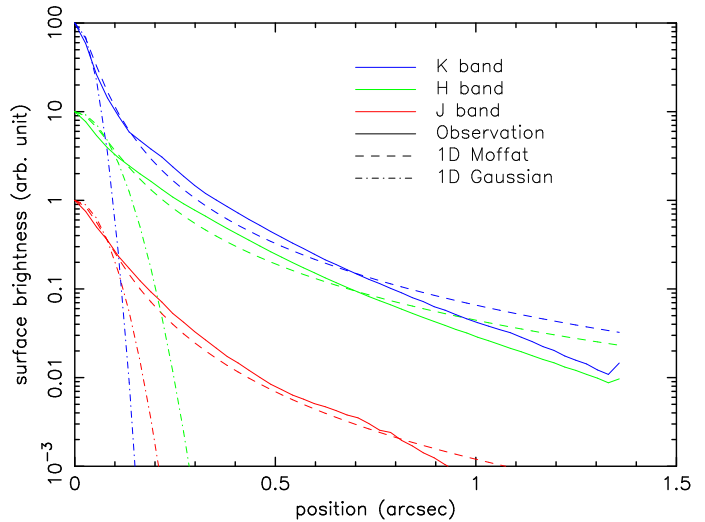}}
  \caption{Radial plot of the PSFs in the $J$ (red), $H$ (green), and $K_S$
           (blue) bands.  The surface brightnesses are normalized.
           The observations are compared with the two one-dimensional PSF
           models of Moffat and Gaussian functions.
         }
  \label{psfmodel}
\end{figure}

\begin{figure*}
  \centering
  \resizebox{\hsize}{!}{\includegraphics{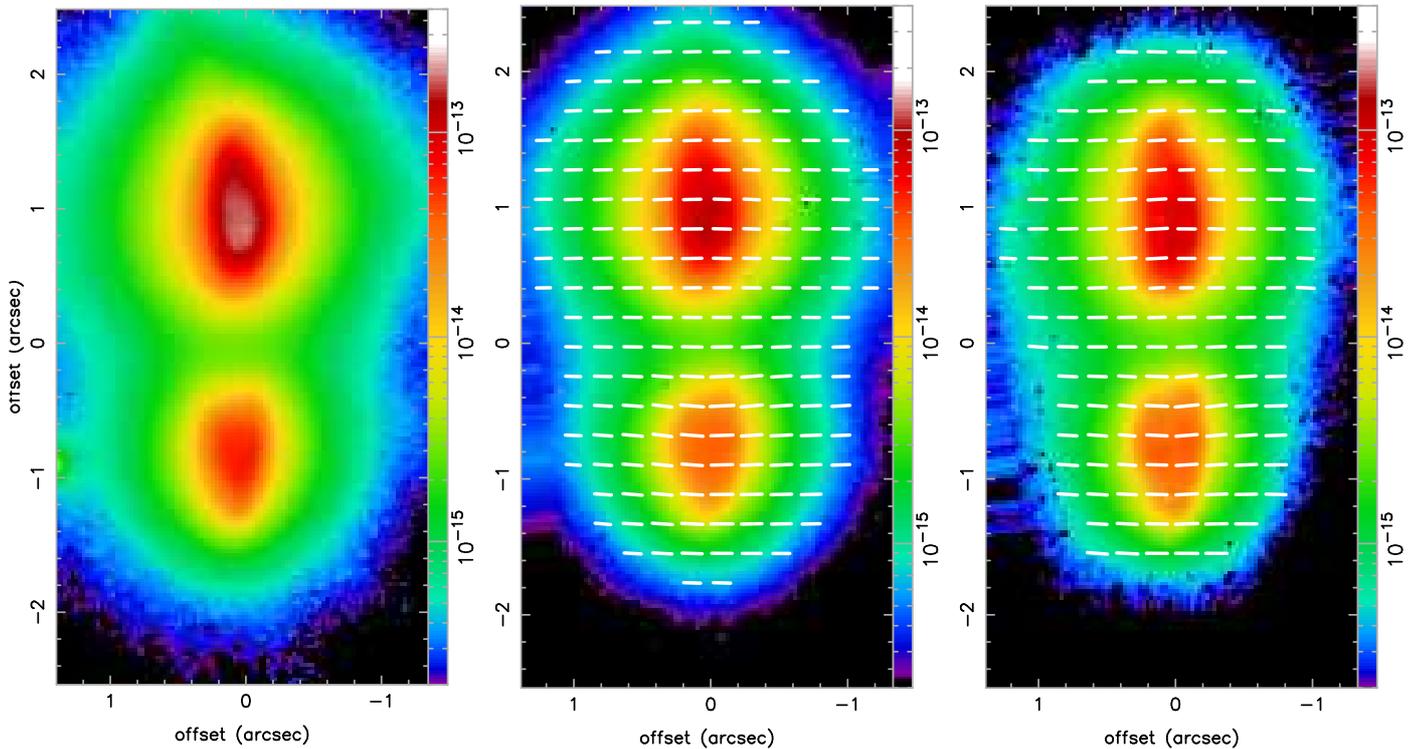}}
  \caption{$J$, $H$, and $K_\mathrm{S}$ band images of IRAS 16342--3814
           obtained using NACO on the VLT.  The polarization vector lines are
           overlaid on the $H$ and $K_S$ band images.  The field of view is
           $2\farcs7\times5\arcsec$.  The images are rotated counterclockwise
           by 107$\degr$.  The color scale bar indicates the surface brightness
           in Wm$^{-2}\mu$m$^{-1}$arcsec$^{-2}$.
         }
  \label{nacopol}
\end{figure*}

We evaluated the shapes and sizes of the beam.  In the $J$ band, a single star
detected in the science frame was examined.  In the $H$ and $K_S$ band the
standard star was used because no single stars are detected in the science
frames.  We found that most of flux from the reference stars is detected
within $1\farcs5$ from the central star, and the star feature is nearly
circular.  Therefore, the star feature was tangentially averaged and was
examined as a one-dimensional PSF.  Figure\,\ref{psfmodel} shows the radial
plot of the PSFs in the $J$, $H$ and $K_S$ bands.  The solid curves are for
the observations.  The measured full width at half maximum (FWHM) are
$0\farcs11$, $0\farcs12$, and $0\farcs07$ in the $J$, $H$ and $K_S$ bands,
respectively.  The real PSF sizes of the object would be somewhat larger than
them because an offset AO referencing was applied.  We examined a Gaussian
model \citep[e.g.][]{bendinelli90} and a Moffat model \citep{moffat69}.
The model PSF forms are given by
$F_\mathrm{Gauss}\left(r\right)=A\exp\left[-1/2\left(r/\sigma\right)^2\right]$
and
$F_\mathrm{Moffat}\left(r\right)=A/\left[1+\left(r/\sigma\right)^2\right]^\beta$,
where $\sigma$ is the beam radius and $\beta$ determines the overall shape of
the PSF.  The resulting parameters and the radial plot of the modeled PSF are
listed in Table\,\ref{psf_parm} and Fig.\,\ref{psfmodel}, respectively.
The Gaussian function is only able to fit around the central star and can
provide a good estimate of the beam size.  On the other hand, the Moffat
function fits better throughout the entire PSF.  Although the resulting model
of this simple analysis would be inaccurate compared to more sophisticated image
reconstruction methods such as the Richardson-Lucy deconvolution, it is useful
for our purpose of comparing images produced by radiative transfer modeling
with the observations.

\subsection{Results}\label{results}
Figure~\ref{nacopol} shows the intensity (Stokes $I$) images in the $J$, $H$,
and $K_S$ bands.  The polarization vector lines are overlaid on the $H$- and
$K_S$ band images.  The intensity images show a peanut-like pair of bipolar
lobes with approximate extensions of $2\farcs5\times4\arcsec$ and the
appearance is nearly constant in all bands.  The western (upper) lobe is about
three times brighter than the eastern (lower) one.  These results are consistent
with Keck II images presented by \citet{sahai05}.  Our images show uneven flux
structures in the bipolar lobes but do not clearly detect the corkscrew-like
structures that are seen in the Keck II $L_p$-band image.

The degree of polarization reaches $P_\mathrm{H}\sim25~\%$ and
$P_\mathrm{K}\sim20~\%$ in the brighter lobe and $P_\mathrm{H}\sim38~\%$ and
$P_\mathrm{K}\sim38~\%$ in the fainter lobe.  The polarization vectors are
aligned along the equatorial plane.  The physical reason is the PSF smoothing
effect, namely, the highly scattered polarized light in the bipolar lobes
spreads with an extended PSF halo component (see Fig.\,\ref{psfmodel}).
This cancels the polarization component perpendicular to the equatorial
direction and the parallel component remains.  Indeed, our model results
convolved with the modeled PSF reproduce the vector alignment as seen in
Fig.\,\ref{img}.

One of the aims of our NIR imaging polarimetry was to detect the
three-dimensional dusty structure in I16342's CDS.   If the corkscrew-like
pattern, as detected in Sahai's $L_p$ band image, is formed by a jet,
the degree of polarization would gradually vary along the spiral because the
scattering angle also varies along it.  Another is the hollow structure in the
bipolar lobe, which could be detected along the rim of the bipolar lobe as
a polarization enhancement for the optically thin case or the opposite for
the optically thick case.  The hollowness is most likely the result of an an
interaction between the fast post-AGB wind and the slow AGB wind, as discussed
in the (G)ISW model \citep[e.g.][]{kwok82,balick87,icke88,sl89}.  Although an
uneven distribution of polarization is detected in our data, we are not able
to provide suitable interpretations that explain these factors.  Therefore,
we used the polarization data only to estimate the particle sizes in the
bipolar lobes, but not to discuss the dust shell structure in the other
sections.

\begin{table}
  \begin{center}
  \caption[]{Parameters of our radiative transfer modeling.}
  \label{model_parm}
  \begin{tabular}{lll}
  \hline
  \hline
  parameters         & values                 & comments$^1$    \\
  \hline
  \multicolumn{3}{c}{central star} \\
  $T_\mathrm{eff}$   & 3000~K                 & $\la3500$$^2$   \\
  $L_\star$          & 6000~$L_{\sun}$        & $^2$            \\
  $D$                & 2.0~kpc                & $^2$            \\
  $R_\star$          & $2.0\times10^{13}$~cm  & calculated      \\
  \hline
  \multicolumn{3}{c}{disk} \\
  $R_\mathrm{in}$    & 20                     & 20 -- 30        \\
  $R_\mathrm{disk}$  & 200~AU                 & 100 -- 200      \\
  $M_\mathrm{disk}$  & 0.01~$M_{\sun}$        & adopted$^3$     \\
  $a_\mathrm{max}$   & 100.0~$\mu$m           & adopted         \\
  $\tau_\mathrm{V}$  & 410                    & calculated      \\
  \hline
  \multicolumn{3}{c}{torus} \\
  $R_\mathrm{torus}$ & 1000~AU                & 700 -- 1000     \\
  $M_\mathrm{torus}$ & 1.0~$M_{\sun}$         & 0.7 -- 1.0$^4$  \\
  $a_\mathrm{max}$   & 10.0~$\mu$m            & adopted         \\
  $\tau_\mathrm{V}$  & 240                    & calculated      \\
  \hline
  \multicolumn{3}{c}{bipolar lobe and AGB shell} \\
  $R_\mathrm{lobe}$  & 5000~AU                & adopted$^5$     \\
  $\beta$            & 3.0                    & adopted         \\
  $\gamma$           & 0.3                    & adopted         \\
  $\epsilon_\mathrm{in}$  & 0.01              & adopted         \\
  $\epsilon_\mathrm{rim}$ & 150               & 100 -- 200      \\
  $R_\mathrm{out}$   & 12\,000~AU             & adopted         \\
  $M_\mathrm{env}$   & 1~$M_{\sun}$           & 1 -- 1.5        \\
  $a_\mathrm{max}$   & 5.0~$\mu$m             & 2.0 -- 5.0      \\
  \hline
  \end{tabular}
  \end{center}
  $^1$ Ranges give the uncertainty of the corresponding model parameters,
  $^2$ \cite{sahai99}, $^3$ assuming an $a_\mathrm{max}=100.0~\mu$m
  dust model, $^4$ assuming an $a_\mathrm{max}=10.0~\mu$m dust model,
  $^5$ based on comparison of the intensity image.
\end{table}

\section{Dust shell model}\label{model}
We have performed radiative transfer calculations of I16342's CDS using
our Monte Carlo code \citep{murakawa08a}.  The SED, and intensity and
polarization images are considered simultaneously, as done before
\citep{murakawa08b,murakawa10a}.  We first tried some test models to obtain
approximated solutions and to understand which parameters affect which results.
Then, many parameter sets (several thousands in total) were examined by SED
fit, which takes less computation time.  From this result, a dozen good
parameter sets were chosen and their images were evaluated.  We finally
selected one that reproduced the shape of the SED, bipolar appearance,
and obscuration of the central star in the optical to NIR, and had nearly
constant NIR polarizations in the bipolar lobe.

Sect.\,\ref{starpar}, \ref{modelgeom}, and \ref{dustmodel} describe the model
assumptions and we compare the model results with the observations in
Sect.\,\ref{modelresult}.

\subsection{Stellar parameters}\label{starpar}
The stellar temperature $T_\mathrm{eff}$ was estimated to be $\sim$15,000~K
from the optical photometry with a $15\arcsec$ aperture \citep{vhg89},
whereas a strong 2.3~$\mu$m CO absorption was detected with a $5\arcsec$
aperture, indicating $T_\mathrm{eff}\la3500~$K.  \cite{sahai99} pointed out
that this strong discrepancy is due to a contamination by the fluxes of two
background stars ($m_V$=14.9 and 13.9~mag) in the $15\arcsec$ aperture, which
are brighter than I16342 ($m_V$=15.6~mag).  Because the central star is
invisible in the optical and infrared, no precise investigation of the stellar
properties have been reported.  We assumed a $T_\mathrm{eff}$ of 3000~K.
The stellar luminosity and the distance are also not well constrained.
Therefore, we adopted a luminosity of 6000~$L_{\sun}$, typical for AGB and
post-AGB stars, and a distance of 2~kpc, as considered by
\citet{sahai99,sahai05}.  The central star, as a radiation source, was assumed
to have a blackbody spectrum.  We find that an effective temperature of 3000~K
fits better than 15\,000~K.

\subsection{Model geometry}\label{modelgeom}
Several geometry forms have been proposed for the CDSs around AGB and post-AGB
stars \citep[e.g.][]{kw85,meixner99,ueta03,oppenheimer05}.  Those proposed
by \cite{kw85}, \cite{meixner99}, and \cite{ueta03} have a latitudinal density
gradient with a higher density at the equatorial region.  Although these
structures reproduce fan-shape morphologies for high equator-to-polar mass
density ratios, e.g.\,$\ga$10, the interaction of the stellar winds and
an optically thick disk are not explicitly expressed.  \citet{oppenheimer05}
used one including an inner disk, a pair of bipolar lobes, and an outer
spherical AGB shell.  We employed a similar form for our previous work of
a PPN M~1--92 where the bipolar appearance was reproduced well with this
geometry \citep{murakawa10a}.

We began with the latter form for I16342.  However, we encountered a problem.
The SED of this object has moderate NIR and strong MIR fluxes.
At first glance, this (semi-) double-peaked flux structure can be explained
with the standard classification of the post-AGB objects
\citep{hrivnak89,kwok93}.  In this evolutionary phase, the intensive mass loss
stops and the dust shell detaches from the central star.  This geometry with
a large inner boundary causes less emission from hot dust at 3 to 10~$\mu$m
and increases the scattered light in the optical-to-NIR.  On the other hand,
I16342 is expected to have a torus that is optically thick even in the MIR
\citep{verhoelst09,lagadec11}.  With the aforementioned geometry form, we could
not find a good solution to fit the SED and to reproduce a bipolar appearance
in the NIR-to-MIR simultaneously.  We attempted a geometry that includes
an additional {\it disk} inside the {\it inner disk}, i.e.\,the inner region
consists of an inner disk and outer torus.  The mass density form of the entire
dust shell is given by
\begin{eqnarray}
  \rho&=&\rho_\mathrm{disk}+\rho_\mathrm{torus}+\rho_\mathrm{lobe}+\rho_\mathrm{AGB},\nonumber\\
  \rho_\mathrm{disk}\left(r,z\right)&=&
    \rho_\mathrm{d}\left(r/R_\mathrm{disk}\right)^{-2.25}
    \exp\left(-\frac{z^2}{2h^2\left(r\right)}\right),\\
  h\left(r\right)&=&
    0.1R_\mathrm{disk}\left(r/R_\mathrm{disk}\right)^{1.25},\\
  \rho_\mathrm{torus}\left(R\right)&=&
    \rho_\mathrm{t}\left(r/R_\mathrm{disk}\right)^{-1.8},\\
  \rho_\mathrm{lobe}\left(R,\theta\right)&=&
    \rho_\mathrm{e}\left(R/R_\mathrm{lobe}\right)^{-2}\times
    \left\{
      \begin{array}{@{\,}l}
        \epsilon_\mathrm{in},\\
        \epsilon_\mathrm{rim},\\
      \end{array}
    \right.\\
  \rho_\mathrm{AGB}\left(R,\theta\right)&=&
    \rho_\mathrm{e}\left(R/R_\mathrm{lobe}\right)^{-2},
\end{eqnarray}
where the coordinates $\left(R,\theta\right)$ and
$\left(r,z\right)=\left(R\sin\theta,R\cos\theta\right)$ are the two-dimensional
spherical and cylindrical coordinates.  The disk and torus are placed in the
regions where $R_\mathrm{in}\le R\le R_\mathrm{disk}$ and
$R_\mathrm{disk}\le R\le R_\mathrm{torus}$, respectively.  The most plausible
definition of a disk should be a region where the matter is gravitationally
bounded by the central star and has a Keplerian rotating motion.
Although dusty disk structures have been spatially resolved in several evolved
stars \cite[e.g.][]{matsuura06,lykou11}, the dynamics are not well known or
much less understood than accretion disks of young stellar objects.  Therefore,
we regarded the disk simply as a geometrically thin, optically thick dust
structure.  In Sect.\,\ref{modelresult} we show that such an inner disk is
required in I16342.

The outer envelope consists of the bipolar lobe and spherical AGB shell.
For the bipolar lobe, the regions
$R_\mathrm{torus}\le R\le R_\mathrm{lobe}\left(\Gamma\left(\theta\right)-\gamma\right)$
and
$R_\mathrm{lobe}\left(\Gamma\left(\theta\right)-\gamma\right)\le R\le R_\mathrm{lobe}$
are the inner cavity and the rim of the lobe, respectively, where
$\Gamma\left(\theta\right)=\left|\theta/\pi-1/2\right|^\beta$ determines the
lobe shape \citep[see][]{oppenheimer05}.  The AGB shell is placed outside
the bipolar lobe with an outer boundary of $R_\mathrm{out}$.  The mass density
coefficients are determined by the masses of the corresponding components.
The dust mass is converted by assuming a gas-to-dust mass ratio of 200.

\begin{figure}
  \resizebox{\hsize}{!}{\includegraphics{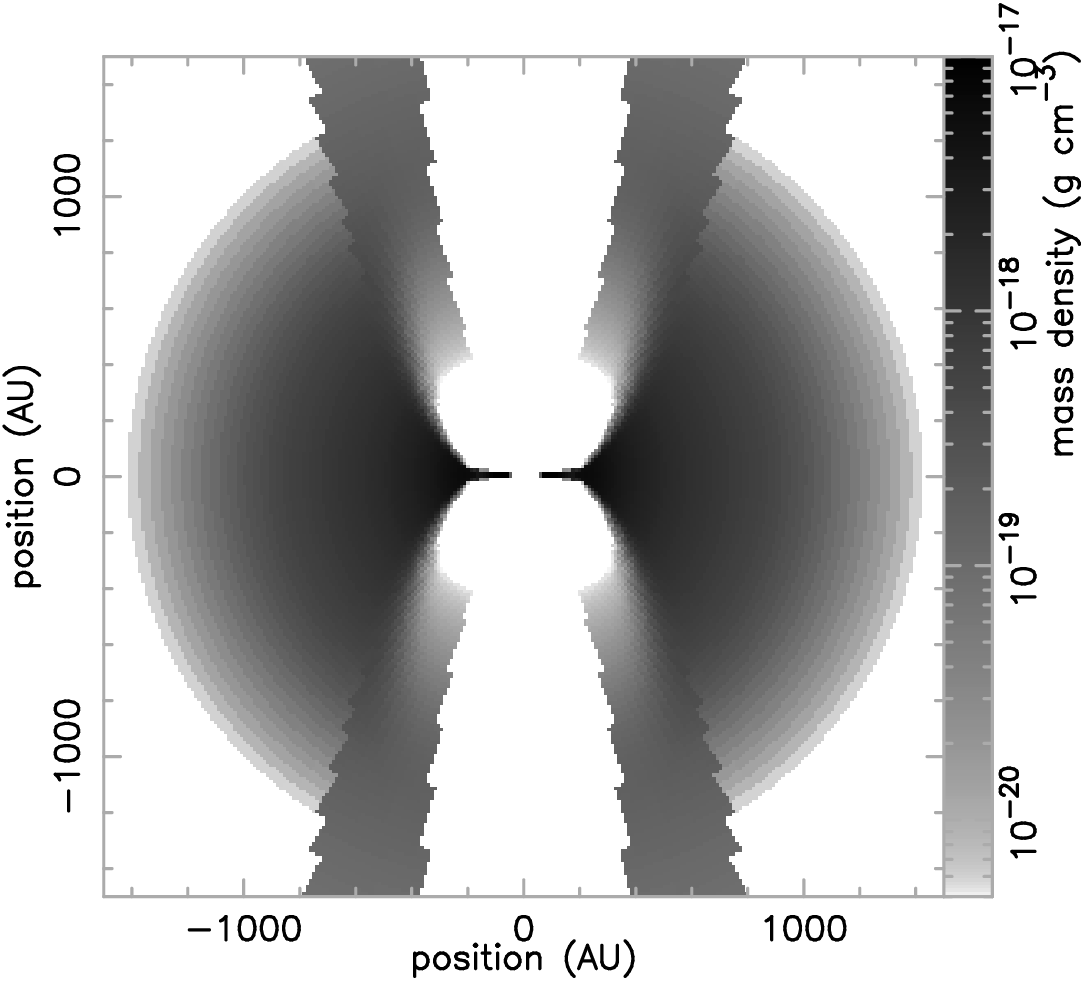}}
  \caption{Mass density map showing the inner 3000~AU region in a plane cutting
           through the symmetry axis.
         }
  \label{rho}
\end{figure}

Figure\,\ref{rho} shows the mass density map in the inner 3000$\times$3000~AU
region. Our modeled geometry is complicated and it is in essence impossible to
determine all parameters accurately from our observational data.  Therefore,
some parameters are set to be fixed values.  Important parameters such as the
inner disk radius $R_\mathrm{disk}$, the disk mass $M_\mathrm{disk}$, the torus
radius $R_\mathrm{torus}$, and the torus mass $M_\mathrm{torus}$ are estimated
from our radiative transfer calculations.

\subsection{Dust particles}\label{dustmodel}
\cite{dwk03} discussed the mineralogy of I16342 based on their ISO spectra
obtained using the {\it Short Wavelength Spectrometer} (SWS).  Their data show
a variety of absorption and emission features attributed to amorphous and
crystalline silicate, frostetite, crystalline water ice, etc.  Investigating
the dust chemistry is one of the interesting and important tasks in dust
formation of evolved stars.  However, a thorough treatment of so many different
chemical compositions and identification of the spatial distributions of
individual species are very complicated tasks and are, therefore, beyond the
scope of this paper.  In our modeling, we adopted simple dust models.  We used
the optical constants for oxygen-deficient silicate \citep{ossenkopf92}.

\begin{figure*}
  \sidecaption
  \includegraphics[width=12.5cm]{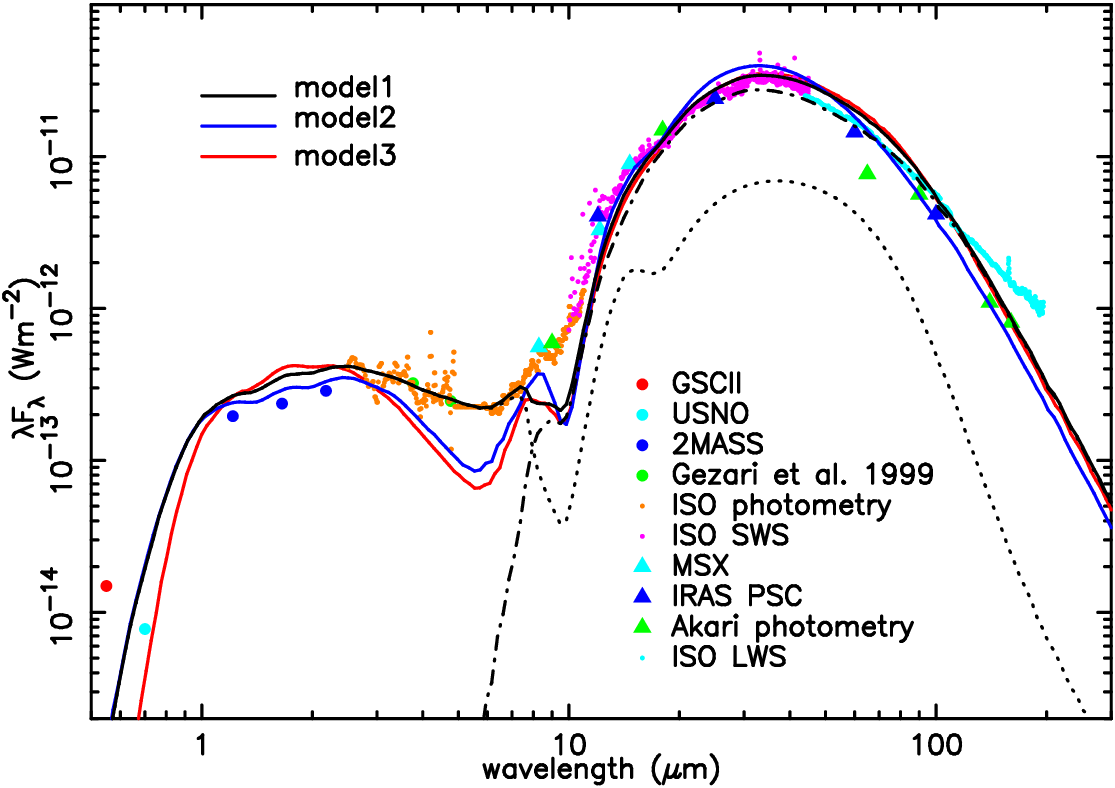}
  \caption{Comparison of SEDs.
           Black lines are the result of the model with an inner disk.
           The solid, dotted, and dash-dot lines denote the total flux,
           the scattered light, and the thermal emission, respectively.
           Red and blue lines denote the results of the same geometry form but
           the disk mass of 0 and a model with only a torus with an inner
           radius of 500~AU.  Colored dots are collected from previous
           observations: US Naval Observatory (USNO), the HST Guide Star
           Catalog (GSC)-II, 2MASS all-sky catalog of point source,
           Catalog of Infrared Observations, Edition 5 \citep{gezari99},
           ISO photometry, SWS and LWS, MSX infrared point source catalog,
           IRAS point source catalog, and Akari$/$IRC mid-IR all-sky Survey
           \citep{ishihara10}
         }
  \label{sed}
\end{figure*}

\begin{figure*}
  \resizebox{\hsize}{!}{\includegraphics{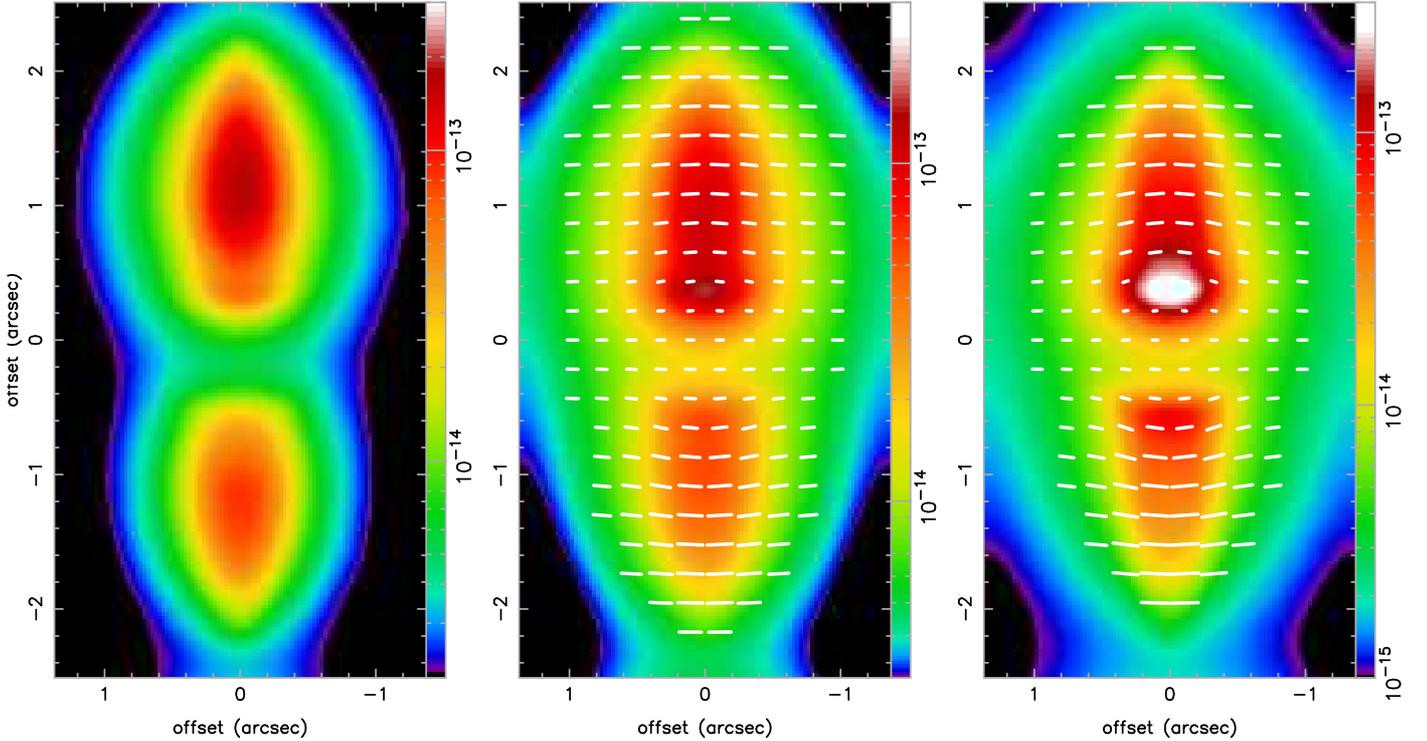}}
  \caption{Results of our two-dimensional radiative transfer calculations.
           The wavelengths are the 1.25~$\mu$m ($J$-band), 1.65~$\mu$m
           ($H$-band), and 2.05~$\mu$m ($K_S$-band) from the left panels.
           The color scale bar indicate the surface brightness in
           Wm$^{-2}\mu$m$^{-1}$arcsec$^{-2}$.  The Moffat PSF model images
           are convolved (see Sect.\,\ref{reduction}).
         }
  \label{img}
\end{figure*}

We assumed different particle sizes for the outer envelope, the torus, and the
inner disk, but the same sizes for the bipolar lobes and AGB shell.
The particle size in the outer envelope was estimated with the polarization
data.  Since the polarizations are high and do not change much in the $H$- and
$K_S$-bands, the fraction of small particles is expected to be higher than
that of the interstellar population.  In our separated paper \citep{murakawa12},
we found a similar result in PPN I18276 and that a size distribution function
with a steep power index provides a good estimate: 0.05~$\mu$m~$\le a$ and
$n\left(a\right)\propto a^{-5.5}\exp\left(-a/a_c\right)$, where $a_c$ is
the cut-off size \citep{kmh94}.  In I16342, $a_c=5.0~\mu$m fits the $H$ and
$K_S$ band polarizations.

Although large grains are expected in the disk and torus, we do not have
(sub-)millimeter flux data or their gas masses and therefore we cannot
constrain the sizes.  Therefore, we provisionally adopted a size distribution
of 0.005~$\mu$m~$\le a\le a_\mathrm{max}$~$\mu$m and
$n\left(a\right)\propto a^{-3.5}$ \citep{mrn77}.
$a_\mathrm{max}$ for the disk and torus are set to be 100~$\mu$m and 10~$\mu$m,
respectively.  We will not discuss dust growth in this object with our
results.

\subsection{Results}\label{modelresult}
Figure\,\ref{sed} shows the SEDs.  The black lines are the result of the
selected model (model~1).  The viewing angle was chosen to be $30\degr$
measured from the equatorial plane.  Although this result is slightly different
compared to that (40\degr) estimated by \cite{sahai99}, this discrepancy is not
important in the following discussion.  The model result fits the observation
well from the optical to FIR, except for the 10~$\mu$m silicate absorption
feature.

We justify the presence of a geometrically thin, inner disk.  First, we compared
with a model without an inner disk (model~2), in which $M_\mathrm{disk}$ is set
to be 0 and the other parameters are identical.  The difference is obvious
in the wavelength range between 3~$\mu$m and 10~$\mu$m.  Although the disk
mass in model 1 is only 0.01~$M_{\sun}$, the MIR flux is governed by the
thermal emission from the hot dust in the inner disk.  The other consideration
is a model without an inner disk where we tried to fit the observations as best
we can (model 3).  This is our first attempt in the modeling (see
Sect.\,\ref{modelgeom}).   We find that models with small inner radii (a few
tens AU) show too strong thermal emission at 3~$\mu$m to 10~$\mu$m and that
a model with a large radius of $R_\mathrm{in}=500$~AU and
$M_\mathrm{torus}=2.0~M_{\sun}$ fits best.  However, as Fig.\,\ref{sed} shows,
the flux between 3~$\mu$m to 6~$\mu$m is too low and model 1 fits better.
The interpretation of I16342's inner region is summarized as follows.
An optically and geometrically thick torus should exist, as previous
observations have shown.  The torus is expected to have a large inner radius
(a few hundred AU) and a high mass (1~$M_{\sun}$).  An additional inner
geometrically thin disk is required, which is responsible for the thermal
emission in the 3~$\mu$m to 10~$\mu$m.

Figure\,\ref{img} shows the monochromatic model images at 1.25~$\mu$m
($J$-band), 1.65~$\mu$m ($H$-band), and 2.05~$\mu$m ($K_S$-band) from left to
right, which are simulated by scattering and absorption only.  In these images,
the Moffat PSF model images (see Sect.\,\ref{reduction}) are convolved.
We confirmed that the Gaussian PSF convolution produces a centro-symmetric
polarization pattern in the bipolar lobe, but the vectors are aligned with
the Moffat PSF model.  This is quite understandable because the surface
brightness of a Gaussian function decays much faster than a Moffat function.
The vector alignment in the entire nebula is reproduced by the PSF smoothing
effect.  Compared with the observations, the striking bipolar appearance and
the presence of a dark lane in the equatorial plane are well reproduced.
Although we do not present the results, we also confirmed that our model image
in the 11.85~$\mu$m shows a bipolar appearance, as seen in the results of
\citet{verhoelst09} and \citet{lagadec11}.

\section{Discussion}\label{discussion}
The mechanisms of PN shaping have been discussed in a number of publications
before.  To form a striking bipolarity with an equatorial narrow waist,
the mass loss material should be highly concentrated in the equatorial plane.
\cite{morris87} modeled mass loss in a binary system from the primary red giant.
Depending on the physical conditions, e.g.\,the mass and the separation of
the secondary, a part of the mass loss material from the primary can be
transferred onto the secondary, which forms a circum-companion disk.
A part of accreting matter can be blown along the polar direction as a jet.
Many bipolar PNs and PPNs, presumably I16342 too, are thought to form by the
binary interaction based on this idea.  Both theories and observations have
developed the binary interaction hypothesis.

Silicate carbon stars are objects in which oxygen-rich species exist around
carbon stars.  According to the current interpretation of the dual chemistry,
the silicate dust that was ejected when the central star was an M giant is
stored in a disk around the hypothetical binary companion and then the central
star is transformed into a carbon star through the third dredge-up
\citep{lloyd90}.  \object{V778~Cyg} is an object of this class, which shows
the 10~$\mu$m silicate emission feature, see \cite{yamamura00}.  These authors
found that the dust has sub-micron sizes and temperatures of 300 --
600~K and is located at about 24~AU from the primary.  The present-day
mass-loss rate is estimated to be about 10$^{-8}$~M$_{\sun}$yr$^{-1}$.
They interpret that the mass loss material can be trapped in the
circum-companion disk, but leaves soon after within a few orbital periods
because the disk is optically thin and the dust is blown away by the strong
radiation pressure from the primary.  With these low mass-loss rates, the disks
are unlikely to be responsible for a bipolar appearance.  It is of interest to
mention the results
of the smoothed particle hydrodynamic simulations for high mass-loss rate cases
of $\ga10^{-5}$~M$_{\sun}$yr$^{-1}$ by
\citet[][ hereafter M98 and M99]{mm98,mm99}.
Their simulations assumed a mass-losing AGB primary and a main-sequence
companion ($M_S=0.25$ -- 2~$M_{\sun}$) with orbital separations of 3.6 -- 50~AU.
In M~1, 2, 10, 12, and 15 out of their 18 models, the density ratios of the
equator to the pole $\rho_\mathrm{eq}/\rho_\mathrm{pol}$ exceeds $\sim$10 and
the nebulae are classified as bipolar.  M~1 and 2 are extreme cases, where the
separations of the companion is small (6.3~AU in their models) and ratios of
the accretion rate to the mass-loss rate are as high as $\sim$0.1, yielding
$\dot{\mathrm{M}}_\mathrm{acc}\ga10^{-6}$~M$_{\sun}$yr$^{-1}$.
$\rho_\mathrm{eq}/\rho_\mathrm{pol}$ exceeds 100.  These systems are likely
to be progenitors of bipolar PNs with narrow waists.  The thick winds
sufficiently shield the strong radiation from the primary and the mass loss
matter is stabilized there for a long time \citep{yamamura00,soker00}, as
expected in the \object{Red Rectangle} \citep{jura95,vanwinckel98},
\object{AFGL 2688} \citep[e.g.][]{bn96}, and probably I16342.

We now consider a simple model of AGB mass loss to see the competition of the
radiation and the shielding based on a similar concept to ``equatorial
back-flow'' introduced by \citet{soker00}.  We assumed that a dust particle is
directly irradiated by the radiation from the central star.  The ratio of the
radiation pressure to the gravitational attraction $\beta_\mathrm{dust}$ is
given by $\beta_\mathrm{dust}=L_\star C_\mathrm{rp}/4\pi GM_\star c$, where
$C_\mathrm{rp}$ is the cross section of the radiation pressure per unit mass,
$G$ the gravity constant and $c$ the velocity of light.  Applying the stellar
parameters of $L_\star=6000~L_{\sun}$ and $M_\star=1$~M$_{\sun}$, a dust model
that is derived from the outer envelope in our modeling,
$C_\mathrm{rp}$=6.73$\times$10$^3$~cm$^2$g$^{-1}$, where the stellar flux is
assumed to dominate at $\lambda=1~\mu$m, $\beta_\mathrm{dust}$ is found to
be $\sim$3$\times$10$^3$.  This high value is quite reasonable in AGB winds,
where the dust is blown away by the radiation pressure.  Next, the shielding
by the mass loss ejecta is considered.  We assumed a spherically symmetric
mass loss at a constant $\dot{\mathrm{M}}$ and constant radial expansion
velocity $V_r$.  The mass density has a power law distribution
$\rho\left(r\right)\propto r^{-2}$, where $r$ is the distance from the central
star.  The optical depth between the inner boundary ($r=R_\mathrm{in}$) and
a position at $r=R$ is given by
\begin{eqnarray}
  \tau=\frac{C_\mathrm{ext}\dot{M}}
            {4\pi\mathrm{V}_r\gamma}
       \left(\frac{1}{R_\mathrm{in}}-\frac{1}{R}\right),
  \label{eq:tau}
\end{eqnarray}
where $C_\mathrm{ext}$ is the extinction cross section per unit mass of the
dust and $\gamma$ the gas-to-dust mass ratio of 200.  If the mass-loss rate
reaches 10$^{-5}$~M$_{\sun}$yr$^{-1}$ at an expansion velocity of 15~kms$^{-1}$,
and $R\gg R_\mathrm{in}=5$~AU is assumed, we derive an optical depth
\begin{equation}
  \tau\approx20.3
      \left(\frac{15~\mathrm{kms^{-1}}}{V_e}\right)
      \left(\frac{\dot{M}}{10^{-5}~M_{\sun}\mathrm{yr}^{-1}}\right)
      \left(\frac{5~\mathrm{AU}}{R_\mathrm{in}}\right),
\end{equation}
where $C_\mathrm{ext}=9.10$$\times$10$^3$~cm$^2$g$^{-1}$.  The dust shell is
optically thick.  When dust particles move away, the radiation is shielded
by the dust that is formed by the subsequent mass loss event.  The optical
depth attains $\sim$8.0$\left(=\ln\beta_\mathrm{dust}\right)$ at $R\sim8.3$~AU
and the attenuated $\beta_\mathrm{dust}$ values becomes $\sim$1.  The radiation
pressure force cannot govern the motion of the particles.  If the system has
a close binary, the gravitational attraction of the secondary can affect the
motion of the mass loss ejecta and can concentrate the flow toward the orbital
plane.  This leads to a flatter geometry, i.e.\,disk.

We have a rough scenario of the inner part of I16342's CDS as follows.
The inner disk of I16342 in our model, which is introduced to be responsible
for the MIR flux, is a circum-companion or circum-binary disk that is formed
by this mechanism.  Because the optical depth of the disk is so high at
$\sim$400 in the $V$-band that the effect of the radiation is negligible,
the disk can be gravitationally bounded by the central star(s).  On the other
hand, the torus is formed in a somewhat different way.  If we assumed that
I16342 has a binary companion, an angular momentum that is supplied by a close
companion (say M$_\mathrm{S}\sim1$~M$_{\sun}$) is insufficient for the entire
torus to have a keplerian rotating motion.  Therefore, the torus is probably
a structure that is accumulated from dust that is escaping from the central
star(s).

\section{Conclusion}\label{conclusion}
We modeled the CDS of the WFS IRAS 16342--3814 with two-dimensional radiative
transfer calculations.  The model parameters were constrained by fitting the
SED collected from various sources and the polarization images obtained using 
the VLT$/$NACO instrument.  I16342 shows a striking bipolar appearance from
the optical to the MIR wavelength ranges.  Taking into account a large viewing
angle of $30\degr$ -- $40\degr$, an optically thick dust torus structure is
expected in the equatorial plane.  We first tried to fit a model consisting of
a geometrically thick torus and an outer envelope with bipolar lobes and
a spherical AGB shell.  However, these models cannot reproduce the SED and
the bipolar appearance simultaneously.  We found that an additional component
of an optically thick, geometrically thin disk is required inside the torus.
In our modeling, different dust sizes were assumed for the disk, the torus, and
the outer envelope.  The particle sizes in the outer envelope were estimated
from the polarization data.  The polarization reaches 25 -- 38~\% in the $H$
band and 20 -- 38~\% in the $K_S$ band in our observations.  This weak
wavelength dependence is explained with a steeper power index of $-5.5$ of
the size distribution function than that of $-3.5$ for the ISM.  Similar
results are found in some other PPNs.  Because of the submillimeter and
millimeter flux information and the estimation of the disk are missing, the
particle sizes of these component cannot be constrained well.  We adopted dust
models of $a_\mathrm{max}=100~\mu$m for the disk and $a_\mathrm{max}=10~\mu$m
for the torus.  With this assumption, the masses of these components were
determined to be 0.01~$M_{\sun}$ and 1~$M_{\sun}$, respectively, but have large
uncertainties.

We discussed a possible formation scenario for the disk and torus.
The important factors are the mass-loss rate and the binary interaction.
For low- to intermediate-rates, the system is likely to have an optically thin
circum-companion disk.  To form an optically thick disk, the mass-loss rate
would need to be so high that the thick wind blocks the radiation from
the primary.  This allows a part of the mass loss ejecta to be condensed in
the equatorial region and to stay there for a long time.  The remains, which
are leaving from the central star(s), form the torus.

\section*{Acknowledgments}
The NIR polarimetric images were obtained at the European Southern Observatory
(proposal ID: 383.D-0197).  The photometric and spectroscopic data are data
produces from the Two Micron All Sky Survey, which is a joint project
of the University of Massachusetts and the Infrared Processing and Analysis
Center$/$California Institute of Technology, funded by the National Aeronautics
and Space Administration and the National Science Foundation, AKARI, a JAXA
project with the participation of ESA, Infrared Space Observatory, and
Infrared Astronomical Satellite.  We thank Luke T. Maud for his encouraging
discussions and proofreading of the paper.


\end{document}